\documentclass{jaa}

\usepackage{graphicx}

\newcommand{\psrj}{PSR\,J0218+4232}

\begin{document}

\title{A new look at distances and velocities of neutron stars}


\author{Frank Verbunt\textsuperscript{*} \and Eric Cator}
\affilOne{Institute of Mathematics, Astrophysics and Particle Physics,
Radboud University, PO Box 9010, 9500 AL Nijmegen, the Netherlands}


\twocolumn[{

\maketitle

\corres{F.Verbunt@astro.ru.nl}

\msinfo{submitted}{revised}{accepted}

\begin{abstract}
We take a fresh look at the determination of distances and velocities
of neutron stars. The conversion of a parallax measurement into
a distance, or distance probability distribution, has led to a debate
quite similar to the one involving Cepheids, centering on the question
whether priors can be used when discussing a single system. 
With the example of \psrj\ we show that a prior is necessary to
determine the probability distribution for the distance.
The distance of this pulsar implies a gamma-ray luminosity larger than 10\%\ of
its spindown luminosity.
For velocities the debate is whether a single Maxwellian describes
the distribution for young pulsars. By limiting our discussion to accurate (VLBI)
measurements we argue that a description with two Maxwellians,
with distribution parameters $\sigma_1=77$ and $\sigma_2=320$\,km/s,
is significantly better. Corrections for galactic rotation, to derive
velocities with respect to the local standards of rest, are insignificant.
\end{abstract}


}]


\doinum{12.3456/s78910-011-012-3}
\artcitid{\#\#\#\#}
\volnum{123}
\year{2016}
\pgrange{1--10}
\setcounter{page}{1}
\lp{25}

\section{Introduction}

This paper summarizes some of the results of a new look at pulsar
distances (Igoshev et al.\ 2016) and pulsar velocities (Verbunt et
al.\ 2017). We add some explanation and some illustrative computations.

The determination of distances to neutron stars is important because it
forms the basis of the determination of their spatial density, and
through this of their birth rate. This in turn has consequences for
our ideas about the progenitors of neutron stars, in particular for the
question of the lowest possible mass for a neutron star
progenitor (e.g.\ Blaauw 1985, Hartman et al.\ 1997).  Because of this
importance, various indirect methods have been developed to establish
distances, in addition to the direct geometric method of parallax
measurement.  In Section\,\ref{s:par} we compare the frequentist and
Bayesian approaches to the determination of distance from a parallax
measurement, to show that priors contribute significantly to the
accuracy of the analysis. 

In Section\,\ref{s:dm} we take a brief look at a method for the distance
determination that uses the dispersion measure and the luminosity
function (cf.\ Verbiest et al.\ 2012). To derive a distance from the
dispersion measure requires a model for the galactic electron-density
distribution, and its accuracy depends critically on this model. It
follows that the method should be used with care, as underestimation
of errors may directly affect the conclusions drawn.  In
Section\,\ref{s:timing} we compare the proper motions determined from
timing with those determined from VLBI interferometry.  In our
description of the velocity distribution of
young pulsars, we limit ourselves to pulsars for which distance and
proper motion are derived from accurate VLBI measurements (Section\,\ref{s:vel}). We briefly
discuss simple indications that the previously derived distribution,
approximated by a Maxwellian with distribution parameter
$\sigma\simeq265$\,km/s (Hobbs et al.\ 2005) is not acceptable. We then apply 
a full analysis to show that a description
with the sum of two Maxwellians does better justice to the observation
of a relatively large number of pulsars with low velocities (Section\,\ref{s:maxwell}).

\section{Distance from parallax \label{s:par}}

Faucher-Gigu\`ere \&\ Kaspi (2006), in their investigation of the
birth velocity of pulsars, give an equation (their Eq.2) that converts
the uncertainty of the parallax measurement into the uncertainty of
the distance.  This equation is in serious error, as a result of
confusion between the frequentist and Bayesian approaches to the
treatment of measurement errors. (We explain this in more detail
below, in Sect.\,\ref{s:wrong}.) A similar error is made (their Eq.3) in the conversion of
uncertainty in the dispersion measure to the uncertainty of the
distance (as detailed below, in Sect.\,\ref{s:dm}). 
Unfortunately, these errors have been repeated in several
later papers by Verbiest et al.\ (2010, 2012, 2014).

Incidentally, the confusion between the frequentist and Bayesian
approaches is also in evidence in the study of Cepheid distances, in a
slightly different form.
Several authors, even in fairly recent papers, state that the parallax of a
single object is not biased (e.g.\ Feast 2002, Francis 2014).
This is all the more surprising as the correct treatment is well
known, as explained in a.o.\ Brown et al.\ (1997),
Sandage \&\ Saha (2002), and  more recently Bailer-Jones (2015).

Brown et al. (1997) also point out that the Lutz-Kelker effect
(Lutz \&\ Kelker 1973)
must be applied with care. In its original form, this effect
is computed with the assumption that the sources are  distributed
homogeneously throughout  space, leading to an {\em a priori} probability
of distance increasing with the square of the distance.
For galactic sources, this assumption does not apply, and a distance
prior must be constructed for each class of objects separately.
In the case of pulsars, Verbiest et al.\ (2012) determine an appropriate
prior for the distances.

\begin{figure}[]
\includegraphics[width=\columnwidth]{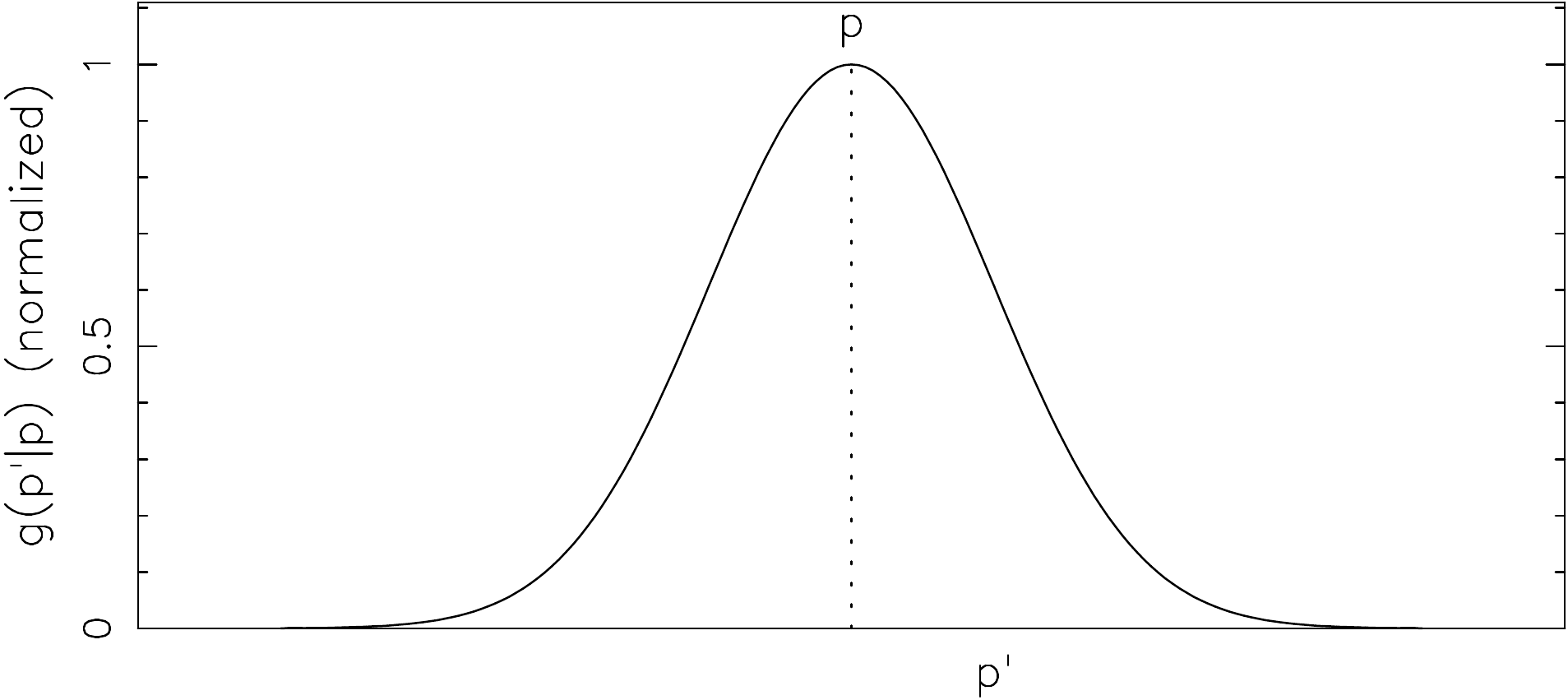}
\caption{The probability of measuring $p'$ when the actual value is
  $p$ for the gaussian case with measurement error $\sigma$.
}\label{f:gauss}
\end{figure}
\begin{figure}[]
\includegraphics[width=\columnwidth]{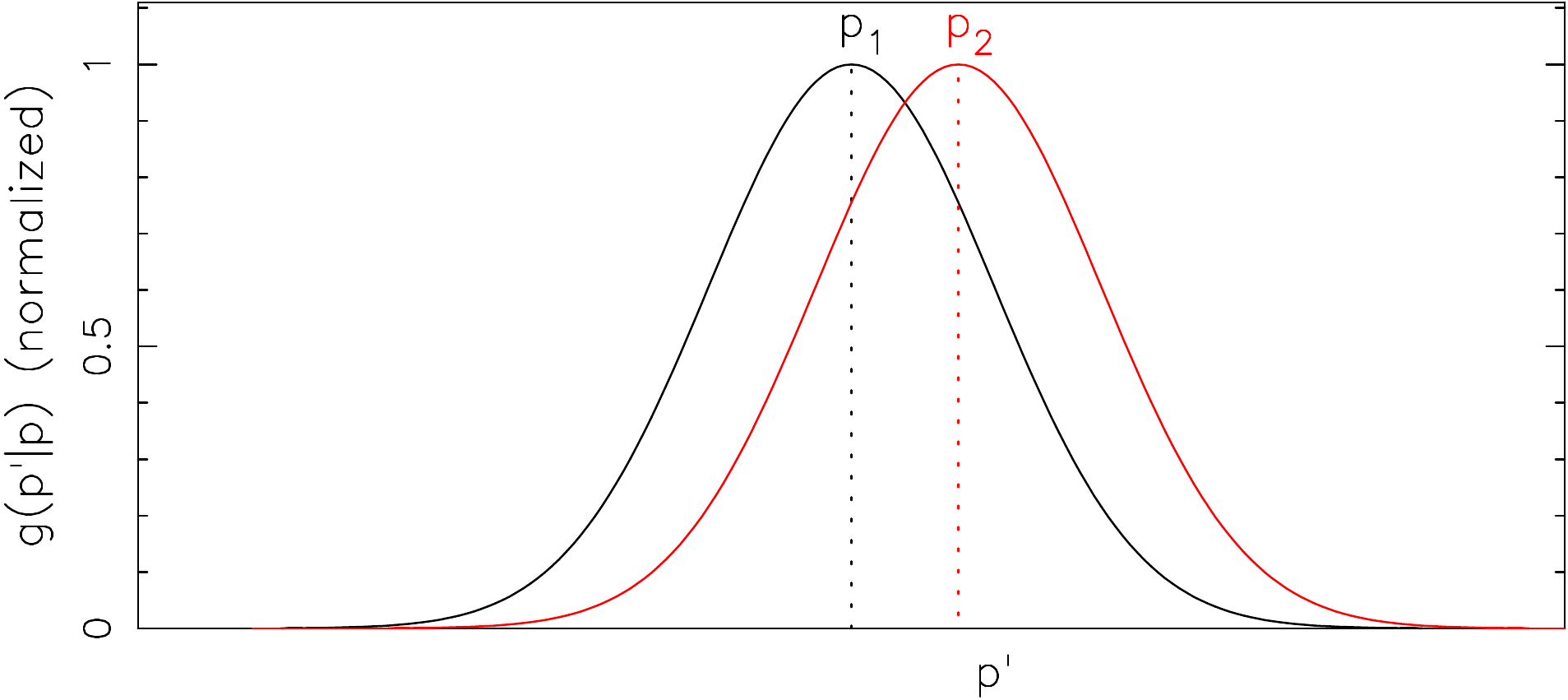}
\caption{The probability of measuring $p'$ when the actual value is
  $p_1$ (black) or $p_2$ (red) for the gaussian case.
}\label{f:twogauss}
\end{figure}

\subsection{Frequentist and Bayesian treatment of measurement errors}

Let us for simplicity assume that the probability $g(p'|p)$ of measuring
parallax $p'$ when the actual parallax is $p$ is given by a gaussian
\begin{equation}
g(p'|p)dp' = {1\over\sigma\sqrt{2\pi}}\exp\left({-(p'-p)^2\over2\sigma^2}\right)dp'
\label{e:gauss}\end{equation}
where $\sigma$ indicates the measurement error (Fig.\,\ref{f:gauss}).
For an actual parallax $p$, this implies that in 68\%\ of the cases
$|p'-p|<\sigma$, i.e.\ $p'-\sigma<p<p'+\sigma$.
Now consider the measurements for two different actual parallaxes,
$p_1$ and $p_2$. For each we have
$$ p'-\sigma<p_1<p'+\sigma \qquad (68\%) $$
$$ p'-\sigma<p_2<p'+\sigma \qquad (68\%) $$
The intervals are the same even when $p_1$ and $p_2$ are
different. More generally for any $p_i$.
$$ p'-\sigma<p_i<p'+\sigma \qquad (68\%) $$

Thus we can state that once a value $p'$ has been measured with
measurement error $\sigma$ {\em the probability is 68\%\ for any
  actual value $p_i$ that the actual value lies in the interval from
  $p'-\sigma$ to $p'+\sigma$}. More generally, for each probability we
can determine a corresponding interval for $p_i$.  For example, there
is a 90\%\ probability that $1.45(p'-\sigma)<p_i<1.45(p'+\sigma)$.
Hence the name frequentist for this approach.  {\em However, from the
  measurement alone we have no information on the probability
  distribution within this interval.}

To obtain that information, we must know how many actual objects
there are with $p_1$, $p_2$,\ldots $p_i$, i.e.\ we must know the
distribution $f(p)$ of $p$. After all, a given measrement $p'$ may
result from any of many actual values $p$, according to
Eq.\,\ref{e:gauss}. 
The joint probability $P(p,p')$ of actual value $p$ and measured value $p'$ is
given by
\begin{equation}
P(p,p')dpdp' = f(p)dp\, g(p'|p)dp'
\end{equation}
and the probability $P(p|p')$ of an actual value $p$ in an interval
$\Delta p$ for a measured value $p'$ is found from this by normalizing
over all possibilities:
\begin{equation}
P(p|p')\Delta p = { f(p) g(p'|p)\Delta p\over\int_p  f(p) g(p'|p)dp}
\end{equation}
where the denominator acts as a normalisation constant.
In this Bayesian approach, $f(p)$ is the prior for $p$.

To apply this to distances we rewrite Eq.\,\ref{e:gauss}
in terms of the distance $D=1/p$:
\begin{equation}
g_D(p'|{1\over D})dp' = {1\over\sigma\sqrt{2\pi}}\exp\left({-(p'-1/D)^2\over2\sigma^2}\right)dp'
\label{e:dgauss}\end{equation}
Note that in this equation, $p=1/D$ is fixed, and that the variable is
$p'$. Hence, in converting Eq.\,\ref{e:gauss} into
Eq.\,\ref{e:dgauss} no $dp/dD$ term is warranted.
For the {\em a priori} distance distribution $f_D(D)$, with
$f_D(D)dD=f(p)dp$ (conservation of numbers), we obtain the
probability of actual distance $D$ when parallax $p'$ is measured as
\begin{equation}
P(D|p')\Delta D = { f_D(D) g_D(p'|{1\over D})\Delta D\over\int_p
  f_D(D) g_D(p'|{1\over D})dD}
\label{e:pdfgauss}\end{equation}

\begin{figure}[]
\includegraphics[width=\columnwidth]{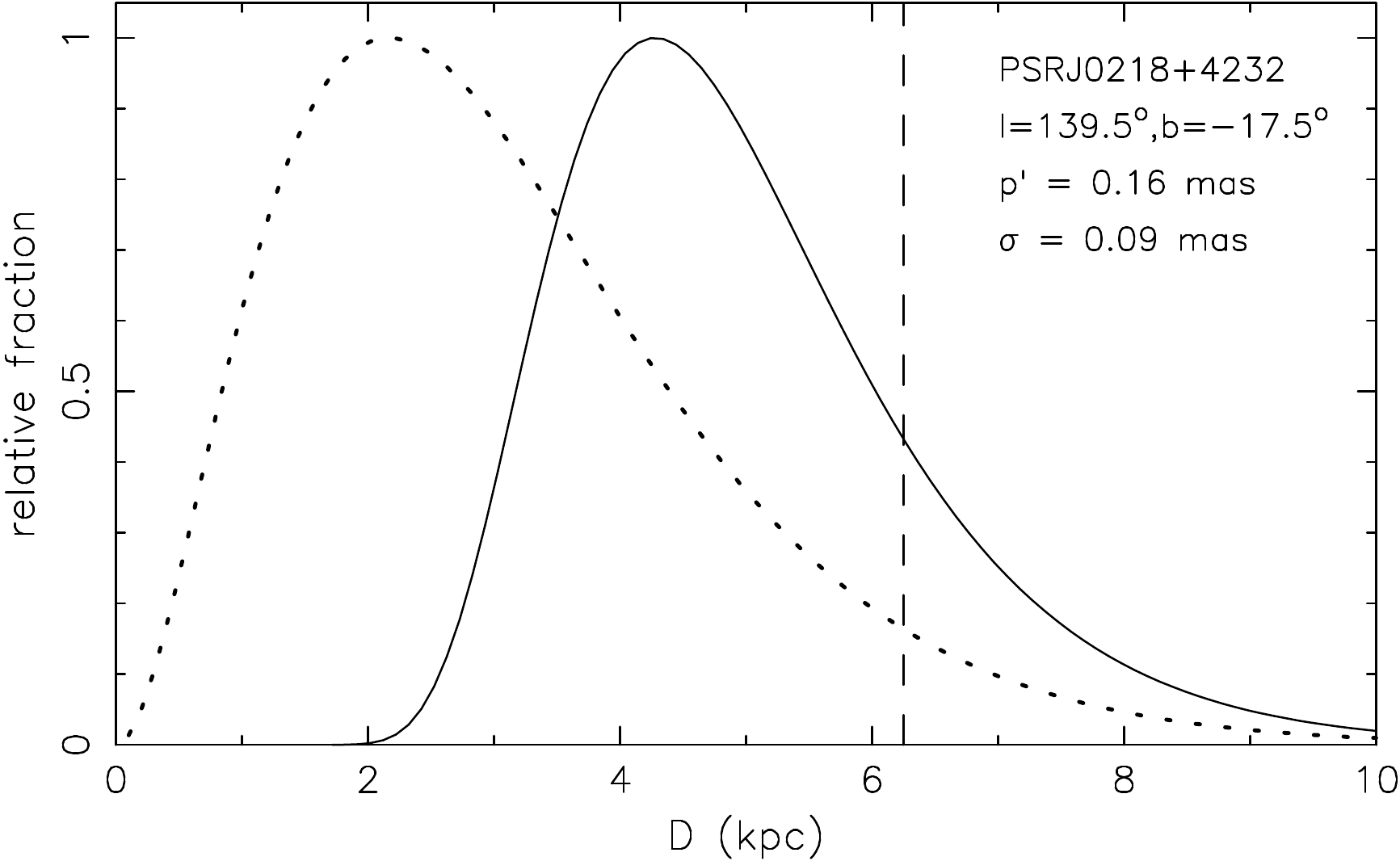}
\caption{The {\em a priori} distribution $f_D(D)$ of millisecond pulsars in the
  direction of \psrj\ (dotted line, Eqs.\,\ref{e:fd}, \ref{e:fdr}), and the measured
  parallax $p'=0.16\pm0.09$\,mas, lead to the distance probability distribution
  given by the solid line, according to Eq.\,\ref{e:pdfgauss}. The
  vertical dashed line indicates the nominal distance $D'=1/p'$.
}\label{f:psrj}
\end{figure}

\subsection{The distance of \psrj}

Igoshev et al.\ (2016)  illustrate this last equation with the case of the millisecond
pulsar \psrj\ (see Fig.\,\ref{f:psrj}).
The distance prior is taken from Verbiest et al.\ (2012), and reflects
the fact that we are looking from a location $R_o=8.5$\,kpc from the
galactic center at a distribution around this center in the radial
direction, and around the galactic plane in the vertical ($z$)
direction. This leads to (in notation slightly altered from that in
Verbiest et al.\ 2012):
\begin{equation}
f_D(D)dD = D^2R^{1.9}\exp\left(-{D\sin b\over0.5\,\mathrm{kpc}} - {R\over1.7\,\mathrm{kpc}}\right)
\label{e:fd}\end{equation}
where $R$ is the distance of the pulsar to the galactic center,
projected on the galactic plane:
\begin{equation}
R = \sqrt{{R_o}^2 + (D\cos b)^2 -2D\cos b\, R_o\cos l}
\label{e:fdr}\end{equation}
This prior is shown in Figure\,\ref{f:psrj} as a dotted line, for the
direction of \psrj.
Eq.\,\ref{e:dgauss} shows that a measured parallax $p'$ can result
from a range of distances;  the probability that a measured $p'$
is due to an actual distance $D$ scales with the product of
Eq.\,\ref{e:dgauss} with the number $f_D(D)$ of objects at that
distance $D$. After normalization this leads to the probability density function 
expressed in Eq.\,\ref{e:pdfgauss}, and shown for \psrj\ in Fig.\,\ref{f:psrj}.

The factor $g_D$ in Eq.\,\ref{e:pdfgauss} leads to a shift of the
most probable value of distance $D$ from the peak of $f_D(D)$ 
to values closer to the distance $D'=1/p'$.
Conversely, the factor of the prior $f_D(D)$ 
leads to a shift of the most probable value of $D$
from the nominal distance $D'$ towards the peak of the prior 
distribution. 

In the basic form of  the Lutz-Kelker effect, for a  homogenous
distribution $f_D(D)\propto D^2$, the most probable actual distance
is always larger than the nominal distance $D'=1/p'$.
Figure\,\ref{f:psrj}  illustrates the fact that the Lutz-Kelker
effect in a more general form, i.e.\ allowing other forms of $f_D(D)$,
may cause the most probable distance to be {\em lower} than
the nominal one.

\subsection{Confusing frequentist and Bayesian approaches\label{s:wrong}}

For a flat prior, $f_D(D)=\mathrm{const}$, Eqs.\,\ref{e:dgauss} and
\ref{e:pdfgauss} simplify to
\begin{eqnarray}
P(D|p')\Delta D &\propto &g_D(p'|{1\over D}) \Delta D \nonumber \\
&=&
{1\over\sigma\sqrt{2\pi}}\exp\left({-(p'-1/D)^2\over2\sigma^2}\right)\Delta
D
\label{e:wrong}\end{eqnarray}
This equation is very similar to Eq.\,\ref{e:dgauss}, but there is a
crucial difference: the probability of Eq.\,\ref{e:dgauss} is
normalized by integrating over $p'$, the probability of
Eq.\,\ref{e:wrong} is normalized by integrating over $D$.
Misreading Eq.\,\ref{e:wrong} as valid for an interval 
$\Delta p'$ leads one to write $\Delta p' =(1/D^2)\Delta D$, 
and thereby add a factor $1/D^2$ to Eq.\,\ref{e:dgauss}.
It appears that this is what Faucher-Gigu\`ere \&\ Kaspi have done.

In fact, as may be seen from Eq.\,\ref{e:pdfgauss}, this corresponds
to assuming a prior $f_D(D)\propto 1/D^2$.

\section{Distance from dispersion measure or luminosity \label{s:dm}}

In principle the dispersion measure $DM$, giving the integrated number of
electrons between Earth and the pulsar, can be combined with a model
for the electron distribution in the Milky Way, to determine the
pulsar distance.
It is well known that this method gives rather uncertain, and
occasionally clearly wrong results (e.g.\ for B1929+10, see 
Table\,5 in Brisken et al. 2002). 
Brisken et al., followed by Faucher-Gigu\`ere \&\ Kaspi (2006)
and by Verbiest (2012), try to circumvent this problem by
`assigning the $DM$ a gaussian probability distribution
function centered on the measured value $DM_o$ with a 40\%\ variance':
\begin{equation}
f_{DM}(DM) \propto \exp\left[-0.5\left({DM-DM_o\over0.4DM_o}\right)^2\right]
\label{e:dm}\end{equation}
This provides a rough guess of the uncertainty of a distance derived
from $DM$ and a model electron dsitribution.

In principle even large measurement uncertainties lead to the correct
result, if the measurements are properly weighted. 
Eq.\,\ref{e:dm} simplifies the complexity of the galactic
electron distribution too much to provide such proper weighting.
Note, for example, that the probability for $DM=0$ (hence $D=0$) is non-zero,
and indeed the same for all values of $DM_o$, no matter how large.
Eq.\,\ref{e:dm} suggests that the error in a distance derived from the
dispersion measure is gaussian, where in fact the error is systematic:
an error in the electron density model leads to a systematic shift in
the derived distance. 

Faucher-Gigu\`ere \&\ Kaspi (2006), followed by Verbiest et al.\
(2012),  compound the error by adding a multiplication factor
$dDM/dD$ in Eq.\,\ref{e:dm}, making an error analogous to the one for
distances discussed in Sect.\,\ref{s:wrong}.  This factor has the
clearly unphysical effect of concentrating the distance probability in
areas of enhanced electron density, since $dDM/dD\propto n_e$.

Verbiest et al.\ (2012) also use the luminosity function to 
constrain the distance: the luminosity function peaks at
low luminosities, hence a pulsar with a given flux is
more likely a nearby low-luminosity one than
a faraway bright pulsar.
In converting a likelihood of luminosity $L$ into a likelihood of
distance, Verbiest et al.\ erroneously introduce a  $d\log L/dD$  
factor.
Igoshev et al.\ (2016) correct this and show that a wide variety
of gamma-ray luminosity functions leads to an isotropic gamma-ray
luminosity in excess of 10\%\ of the spindown luminosity for
PSR\,J0218+4232.
 
Because of the steepness of the luminosity function, straightforward
application of the resulting bias pushes the distance probability
to the lowest distances allowed by other indicators.
Our knowledge of the luminosity function of pulsars
depends on our knowledge of distances, and thus in principle
the luminosity function and distance distribution of pulsars
should be determined together.

\section{Velocity from timing and dispersion measure\label{s:timing}}

The annual variation in the difference between heliocentric and
geocentric pulse arrival times depends on the celestial position of
the pulsar. This dependence may be used to determine the position of
the source, and over time its parallax and proper motion, from pulse
timing. Hobbs et al.\ (2005) list a large number of proper motions for
pulsars determined with this method. By comparing these proper motions
and their uncertainties with the measurements for the same pulsars
obtained with VLBI (by Chatterjee et al.\ 2009, Brisken et al.\ 2002,
Kirsten et al.\ 2015),  we see that the measurement errors
given for young (i.e.\ not recycled)
pulsars are of order a hundred times larger for timing
measurements than for VLBI. Because of these large uncertainties no timing
parallaxes have been determined for young pulsars.

Hobbs et al.\ (2005) therefore use distances estimated from dispersion
measure to convert the proper motions into velocities. Their use of a
non-parametric {\tt clean} algorithm to determine the intrinsic
velocity distribution, has the advantage of obviating the need to
prescribe a parametrized form of this distribution. However,
Hobbs et al.\ note that the result is well described by a
Maxwellian with distribution parameter $\sigma=$ 265 km/s,
and  argue that the low values of velocity perpendicular to
the line of sight observed  for some pulsars are the result of
projection effects.

One of us, F.V., has always found it hard to accept this, for the
following reason. An isotropic Maxwellian may be considered as 
composed of three gaussians, in three mutually perpendicular
directions.  If we choose the line of sight as one
direction, the two remaining directions are in the celestial plane,
and the two gaussians lying in this plane may be combined
to give the distribution of $v_\perp$. The fraction of velocities
in this distribution below any $v_c$ may be written (for derivation see
Appendix, Eq.\,\ref{e:vperpcum})
\begin{equation}
f(v_\perp<v_c) = 1 - e^{{-v_c}^2/(2\sigma^2)}
\label{e:vperp}\end{equation}
Table\,5 of Brisken et al.\ (2002) lists the nine accurate velocities
$v_\perp$  known at the time, and of these two have $v_\perp<40$\,km/s.
For $\sigma=265$\,km/s and $v_c=40$\,km/s, the 
probability for one trial that $v_\perp<v_c$ follows from
Eq.\,\ref{e:vperp} to be about 1\%.
The probability of finding 2 in 9 trials is 0.4\%.
This suggests that the fraction of low velocities is underestimated
by the analysis of Hobbs et al.\ (2005). Remarkably,
this original argument for the velocity study of Verbunt et al. (2017)
study was rather weakened when the accurate proper motion data for 
28 pulsars were collected.
Not a single new one with $v_\perp<40$\,km/s was added! The
probability of finding 2 in 28 trials is about 4\%.

\begin{figure}[t]
\includegraphics[width=\columnwidth]{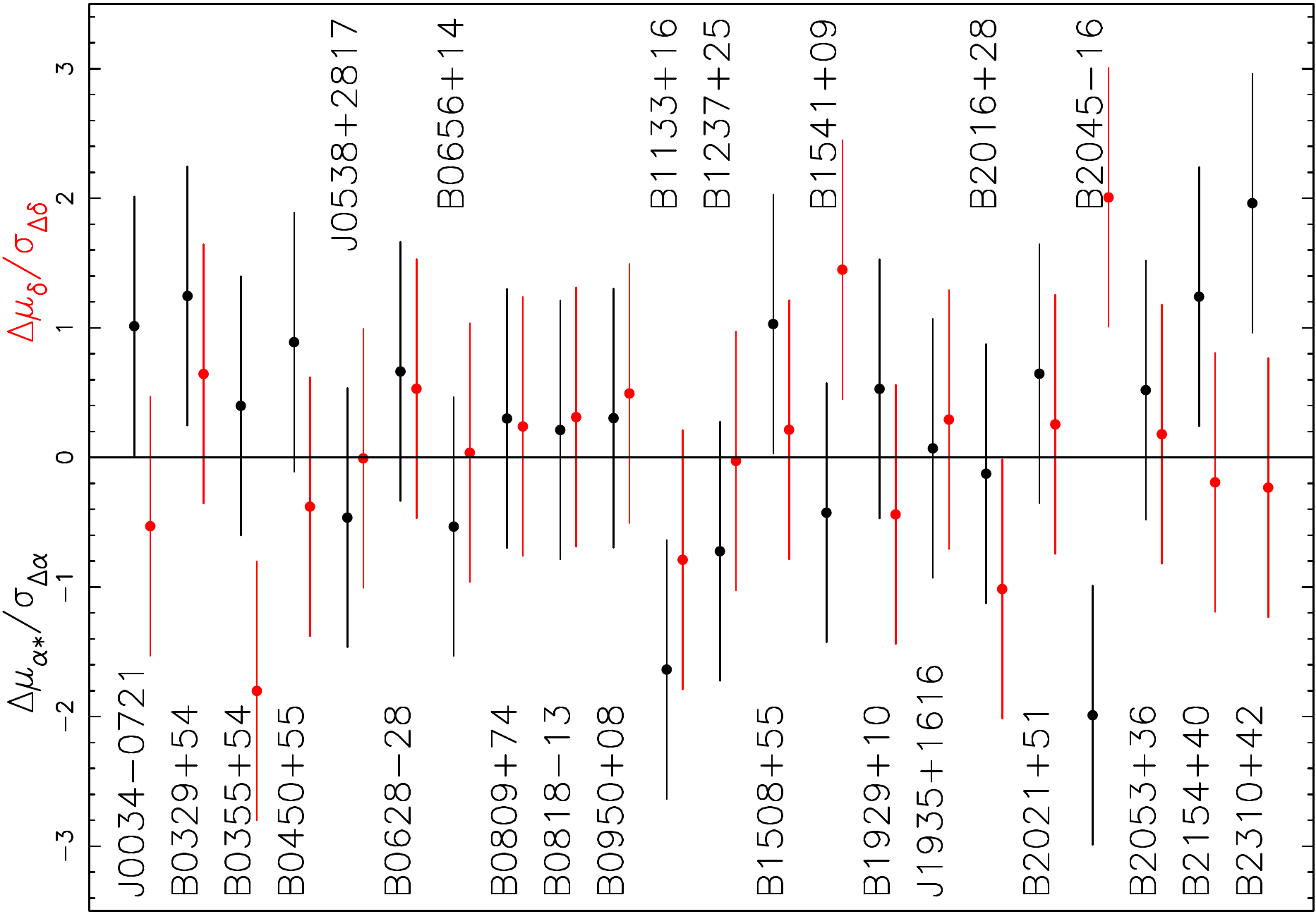}
\caption{The difference  $\Delta\mu_{\alpha*}=\mu_{\alpha*,VLBI}-\mu_{\alpha*,tim}$
between proper motions in the direction of right ascension
$\mu_{\alpha*,VLBI}$ and $\mu_{\alpha*,tim}$ measured with VLBI  and with
timing, respectively, in units of the error in the difference 
$\sigma_{\Delta\alpha}$ (black); and analogous
for the difference $\Delta\mu_\delta$ of the proper motions
in the direction of declination (red). 
}\label{f:difmu}
\end{figure}

As we will see below, a single Maxwellian {\em does} underestimate the
number of low velocity pulsars, albeit at less low velocities than
suggested by the two velocities below 40\,km/s. Such an underestimate
may arise if Hobbs et al.\ (2005) underestimate the velocity errors.

Figure\,\ref{f:difmu} compares the proper motions determined from
timing with those determined from VLBI, for pulsars with accurate VLBI
measurements, by plotting the difference between the proper motions in
units of the error in the difference, for the directions of right
ascension and of declination separately. The Figure shows that the
errors for the timing proper motions, although large, are reliable, in
the sense that they are distributed around the correct (VLBI) values
as expected. Thus, in velocities determined with proper motions
from timing and distances from dispersion measure, the problem
for a reliable statistical analysis lies in the distances.

\begin{figure*}[]
\centerline{\includegraphics[width=15.5cm]{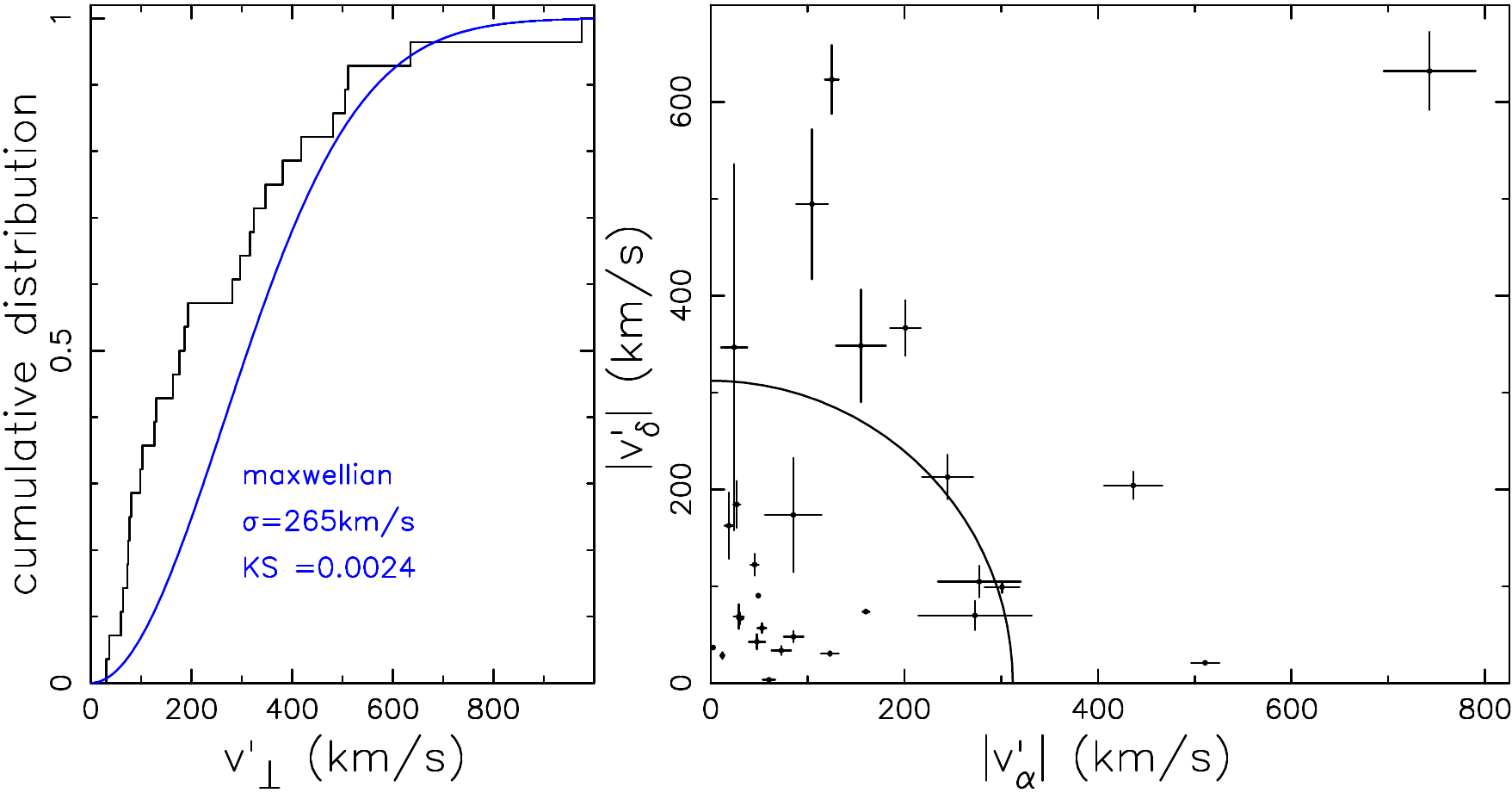}}
\caption{Left: the observed cumulative distribution of $v'_\perp$
  derived from VLBI measurements for 28 pulsars compared with the
  cumulative distribution of $v_\perp$ (Eq.\,\ref{e:vperp}) for a
  Maxwellian with $\sigma=265$\,km/s.  Right: The nominal velocities
  $|v'_\alpha|=|\mu'_{\alpha*}|/p'$ and $|v'_\delta|=|\mu_\delta|/p'$. The circle gives the median
  velocity $v_{\perp,m}$ (Eq.\,\ref{e:median}) for $\sigma=265$\,km/s.}
\label{f:cumul}\end{figure*}

\section{Velocity from VLBI measurements\label{s:vel}}

Given the large errors in the velocities derived with distances
from the dispersion measure and proper motions from timing, it appears
appropriate to make a first effort at determining the velocity
distribution on the basis of the smaller sample with VLBI parallaxes
and proper motions. With these much smaller errors, exact
understanding of the error distribution is less critical.  We collect
from the literature 28 young (in the sense of not recycled) pulsars
for which these data are available.  We indicate the measured values
and the nominal values derived from them with a prime: parallax
$p'$ and proper motions $\mu'_{\alpha*},\mu'_\delta$; and
nominal distance $D'=1/p'$ and velocity perpendicular to the line of sight
$v'_\perp = \sqrt{{\mu'_{\alpha*}}^2+{\mu'_\delta}^2}/p'$.

In Figure\,\ref{f:cumul} we show the cumulative distribution of
$v'_\perp$, together with the cumulative distribution according to
Eq.\,\ref{e:vperp} for $\sigma=265$\,km/s.
The Kolmogorov-Smirnov test gives a probability of 0.0024 that
the observed distribution is drawn from this distribution.
It shows that the Maxwellian predicts too few pulsars with
low velocities, up to several hundred km/s.
Some caution is required in the interpretation of this result, 
because the observed distribution shown in Fig.\,\ref{f:cumul}
and used in the Kolmogorov-Smirnov test, ignores measurement
errors.

In Figure\,\ref{f:cumul} we also show the absolute values
of the nominal velocities $v'_\alpha=\mu'_{\alpha*}/p'$ and 
$v'_\delta=\mu'_\delta/p'$, together
with their nominal errors.  The median of $v_\perp$ is found
by equating the cumulative distribution of Eq.\,\ref{e:vperp}
to 0.5:
\begin{equation}
1 - e^{{-v_{\perp,m}}^2/(2\sigma^2)}= 0.5\Rightarrow v_{\perp,m}=\sigma\sqrt{2\ln2}
\label{e:median}\end{equation}
This median is also shown in Figure\,\ref{f:cumul}. 
It is seen that the errors in the lower velocities are small,
indicating that our conclusion from the Kolmogorov-Smirnov
test on $v'_\perp$ is reliable. Also, only 7 of 28 pulsars
have $v_\perp$ higher than the median velocity predicted
by a Maxwellian with $\sigma=265$\,km/s.

Figure\,\ref{f:cumul} strengthens our earlier suspicion that a single
Maxwellian underpredicts the number of low-velocity pulsars.
For a definite conclusion, however, we must perform an analysis
which takes account of the measurement errors properly.

\section{The interplay of distance, proper motion and velocity
  distribution\label{s:maxwell}}

As a first prior for the intrinsic velocity distribution we consider
a single isotropic Maxwellian. Each pulsar velocity is a draw from this Maxwellian,
i.e.\ a draw from each of three gaussians in mutually perpendicular
directions. For each pulsar, we choose the three directions along the
line of sight and along the directions of increasing right ascension $\alpha$
and declination $\delta$, and thus for the direction along $\alpha$
we have the prior
\begin{equation}
f(v_\alpha,\sigma)dv_\alpha = {1\over\sigma\sqrt{2\pi}} e^{-{v_\alpha}^2/2\sigma^2}dv_\alpha
\label{e:valpha}\end{equation}
and analogously for $v_\delta$ and $v_r$.
The joint probability of measured values for parallax and proper
motions $p'$, $\mu'_{\alpha*}$ and $\mu'_\delta$ and actual distance
and velocities $D$, $v_\alpha$, $v_\delta$  and $v_r$ follows as

\begin{eqnarray}
P_\mathrm{maxw} & \equiv &
P_\mathrm{maxw}(p',\mu'_{\alpha*},\mu'_\delta,D,v_\alpha,v_\delta,v_r)\nonumber \\
&=& f_Df(v_\alpha,\sigma)f(v_\delta,\sigma)f(v_r,\sigma)g_D g_\alpha g_\delta
\label{e:joint}\end{eqnarray}
where $f_D$ is given by Eqs.\,\ref{e:fd}, \ref{e:fdr} and $g_D$ by
Eq.\,\ref{e:dgauss}; $f(v_\alpha,\sigma)$ by Eq.\,\ref{e:valpha}, and
$f(v_\delta,\sigma)$ and $f(v_r,\sigma)$ analogously; and 
$g_\alpha$ and $g_\delta$
by 
\begin{equation}
g_\alpha = {1\over\sigma_\alpha\sqrt{2\pi}}\exp\left[-\,{(\mu_{\alpha*,G}(D)+v_\alpha/D
-\mu'_{\alpha*})^2\over2{\sigma_\alpha}^2}\right] 
\label{e:galpha}\end{equation}
\begin{equation}
g_\delta = {1\over\sigma_\delta\sqrt{2\pi}}\exp\left[-\,{(\mu_{\delta,G}(D)+v_\delta/D
-\mu'_\delta)^2\over2{\sigma_\delta}^2}\right] 
\label{e:gdelta}\end{equation}
where $\sigma_\alpha$ and $\sigma_\delta$ are the
measurement errors in $\mu_{\alpha*}$ and $\mu_\delta$, respectively,
and $\mu_{\alpha*,G}(D)$ and $\mu_{\delta,G}(D)$ the
corrections due to galactic rotation, between the local standards of
rest at the position of the Sun and the pulsar. These
corrections are necessary, because we are interested in the peculiar
velocity of the pulsar, not including the apparent velocity due to
galactic rotation. Because most pulsars with an accurate parallax are
nearby, these corrections generally are small.

To obtain the value of the scale parameter $\sigma$ which gives the
most likely correspondence with the measurements, we must consider
the contributions to the likelihood of all distances and velocities,
i.e.\ integrate Eq.\,\ref{e:joint} over $D$, $v_\alpha$, $v_\delta$
and $v_r$. The integral over $v_r$ is 1; the integrals over $v_\alpha$
and $v_\delta$ are more involved, but can be done analytically.
The resulting likelihood is (Verbunt et al.\ 2017):

\begin{figure}[t]
\includegraphics[width=\columnwidth]{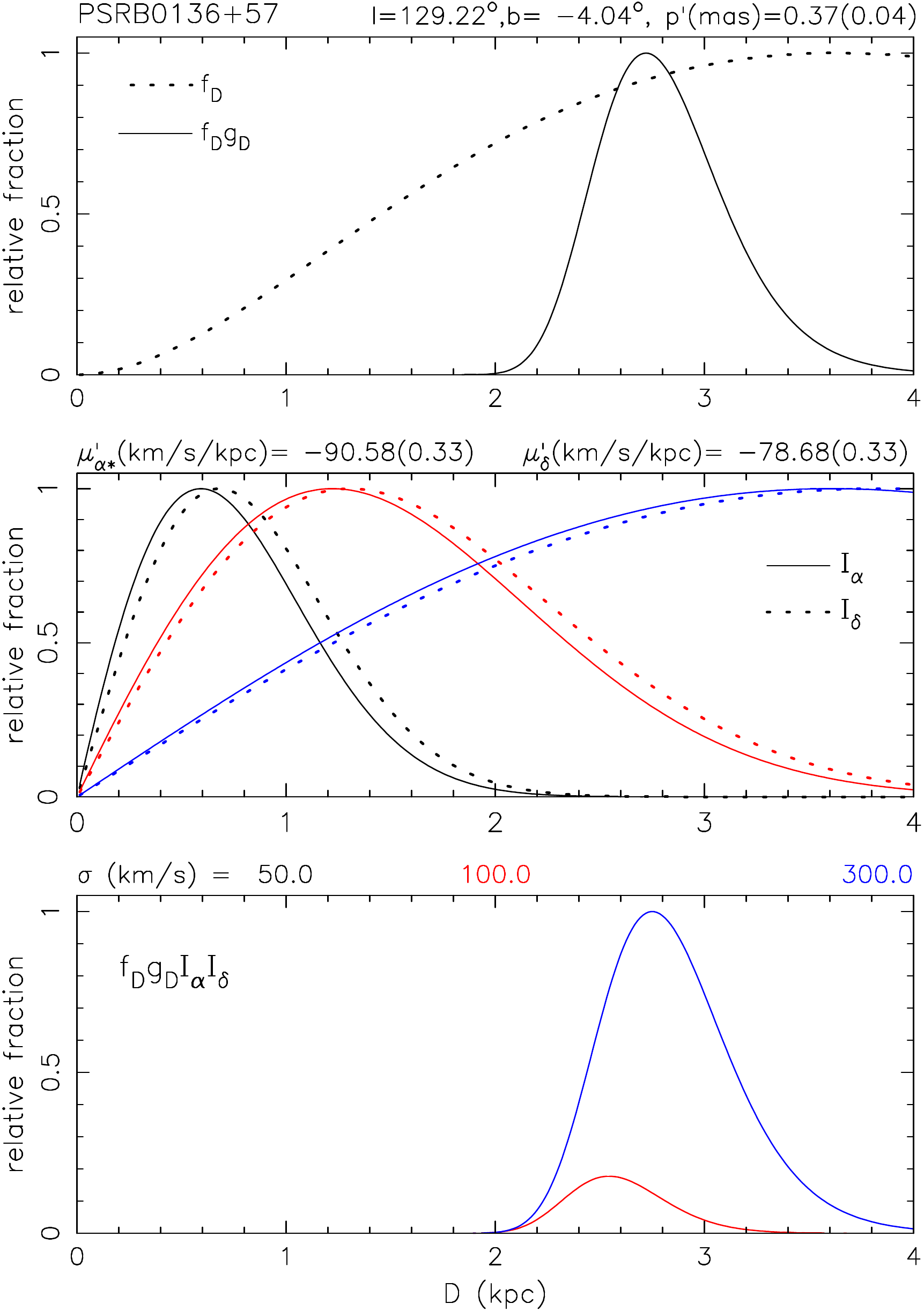}
\caption{The integrand of Eq.\,\ref{e:lmaxw}.
  The distance implied by the parallax and
  galactic pulsar distribution in the direction of PSR\,B0136+57 (top
  frame), combined with the proper motion ($\mathcal{I}_\alpha$,
  $\mathcal{I}_\delta$, Eq.\,\ref{e:ialpha}, middle frame) implies a large velocity
  and favours a Maxwellian with high scale parameter $\sigma$.
  The curves in the top and middle frame are normalized to
  maximum value 1. The lower frame shows the integrand of
  Eq.\,\ref{e:lmaxw}, normalized such that the area under the curve is
  proportional to the likelihood $L_\mathrm{maxw}(\sigma)$
  (Eq.\,\ref{e:lmaxw}).  }\label{f:lmaxw}
\end{figure}

\begin{eqnarray}
L_\mathrm{maxw}(\sigma) &=& \int_0^{D_\mathrm{max}}\int_0^\infty\int_0^\infty\int_0^\infty
 P_\mathrm{maxw}dD dv_\alpha dv_\delta dv_r\nonumber\\
&=& \mathcal{C} \int_0^{D_\mathrm{max}}\, f_Dg_D \mathcal{I}_\alpha \mathcal{I}_\delta dD
\label{e:lmaxw}\end{eqnarray}
where $\mathcal{C}$ is a constant, ${D_\mathrm{max}}$ the maximum
distance (we use ${D_\mathrm{max}}=10$\, kpc; beyond this distance
the factor $g_D$ according to Eq.\,\ref{e:dgauss} ensures that the integrand
is effectively zero for the pulsars in our sample), and we define
\begin{eqnarray}
\mathcal{I}_\alpha & \equiv &\left(1+{\sigma^2\over D^2{\sigma_\alpha}^2}\right)^{-1/2}
\exp\left[-{1\over2}{(D\,\mu_{\alpha*,G}-D\,\mu'_{\alpha*})^2\over
  \sigma^2+D^2{\sigma_\alpha}^2}\right] \nonumber \\
\mathcal{I}_\delta& \equiv &
\left(1+{\sigma^2\over D^2{\sigma_\delta}^2}\right)^{-1/2}
\exp\left[-{1\over2}{(D\,\mu_{\delta,G}-D\,\mu'_{\delta})^2\over
  \sigma^2+D^2{\sigma_\delta}^2}\right] \nonumber \\
 & & \label{e:ialpha}
\end{eqnarray}

The effect of the separate contributors to the integrand of
Eq.\,\ref{e:lmaxw} is shown in Figure\,\ref{f:lmaxw}, for the
case of PSR\,B0136+57. The observational data $p'$, $\mu'_{\alpha*}$ and
$\mu'_\delta$ for this pulsar are taken from Chatterjee et al.\
(2009). We convert the proper motion with
\begin{equation}
\mu(\mathrm{km/s/kpc})=4.74\mu(\mathrm{mas/yr}) 
\end{equation}
The accurate parallax and proper motion imply a velocity of
several hundred km/s: the nominal projected velocity is
$v'_\perp=324$\,km/s. When we compare the probability of such
a velocity for three different Maxwellians, with $\sigma=50$, 100,
and 300\,km/s respectively, the probability of the one with
$\sigma=300$\,km/s is highest.The probability of the Maxwellian
with $\sigma=100$\,km/s is significantly lower, and the Maxwellian with
$\sigma=50$\,km/s is virtually excluded (its integrand invisible
in Figure\,\ref{f:lmaxw}).

\begin{figure}[t]
\includegraphics[width=\columnwidth]{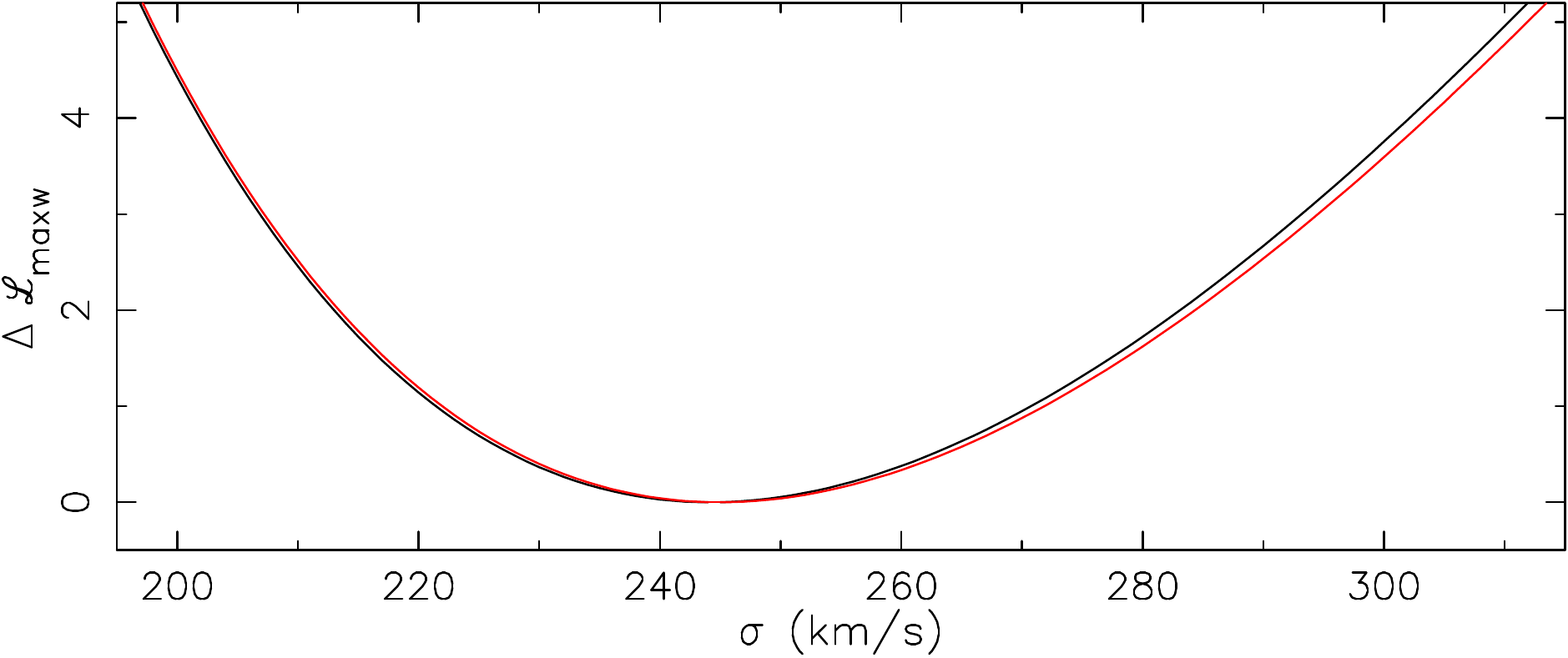}
\caption{$\mathcal{L}_\mathrm{maxw}$ according to Eq.\,\ref{e:total}
 as a function of the Maxwellian scale parameter $\sigma$. The black
 curve shows the result of the full calculation. The red curve, almost
 indistinguishable, shows the result when corrections $\mu_{\alpha*,G}$
 and $\mu_{\delta,G}$ for galactic rotation are omitted.}
\label{f:scriptl}
\end{figure}

\subsection{Description with a single Maxwellian}

To determine the best value of $\sigma$ for the complete set
of 28 pulsars, Verbunt et al.\ (2017) first compute $L_\mathrm{maxw}(\sigma)$ according
to Eq.\,\ref{e:lmaxw} for each of them, integrating numerically over
$D$. From these likelihoods the deviance is computed as 
\begin{equation}
\mathcal{L}_\mathrm{maxw} (\sigma) = -2\sum_{i=1}^N
\ln L_{\mathrm{maxw},i} (\sigma)
\label{e:total}
\end{equation}
where index $i$ labels the pulsar. With this definition of the deviance,
the best value $\sigma_\mathrm{opt}$ is the one that minimizes
$\mathcal{L}_\mathrm{maxw}$ (and thus maximizes the product
of the likelihoods), and the differences 
\begin{equation}
\Delta\mathcal{L}_\mathrm{maxw}\equiv \mathcal{L}_\mathrm{maxw}(\sigma) -
\mathcal{L}_\mathrm{maxw}(\sigma_\mathrm{opt})
\label{e:callik}\end{equation}
approximate a $\chi^2$ distribution.  $\Delta\mathcal{L}_\mathrm{maxw}$
is shown as a function of $\sigma$ in Fig.\,\ref{f:scriptl}.
The minimum of $\mathcal{L}_\mathrm{maxw}$ occurs at
$\sigma_\mathrm{opt}\simeq 245$\,km/s.

To see the effect of the corrections for galactic rotation to the
observed proper motion, we also perform a calculation in which these
corrections are omitted, i.e.\ in which $\mu_{\alpha*,G}$ and $\mu_{\delta,G}$
in Eqs.\,\ref{e:lmaxw} and \ref{e:ialpha} are put to zero. The result
is the same within the uncertainty.

\begin{figure}[t]
\includegraphics[width=\columnwidth]{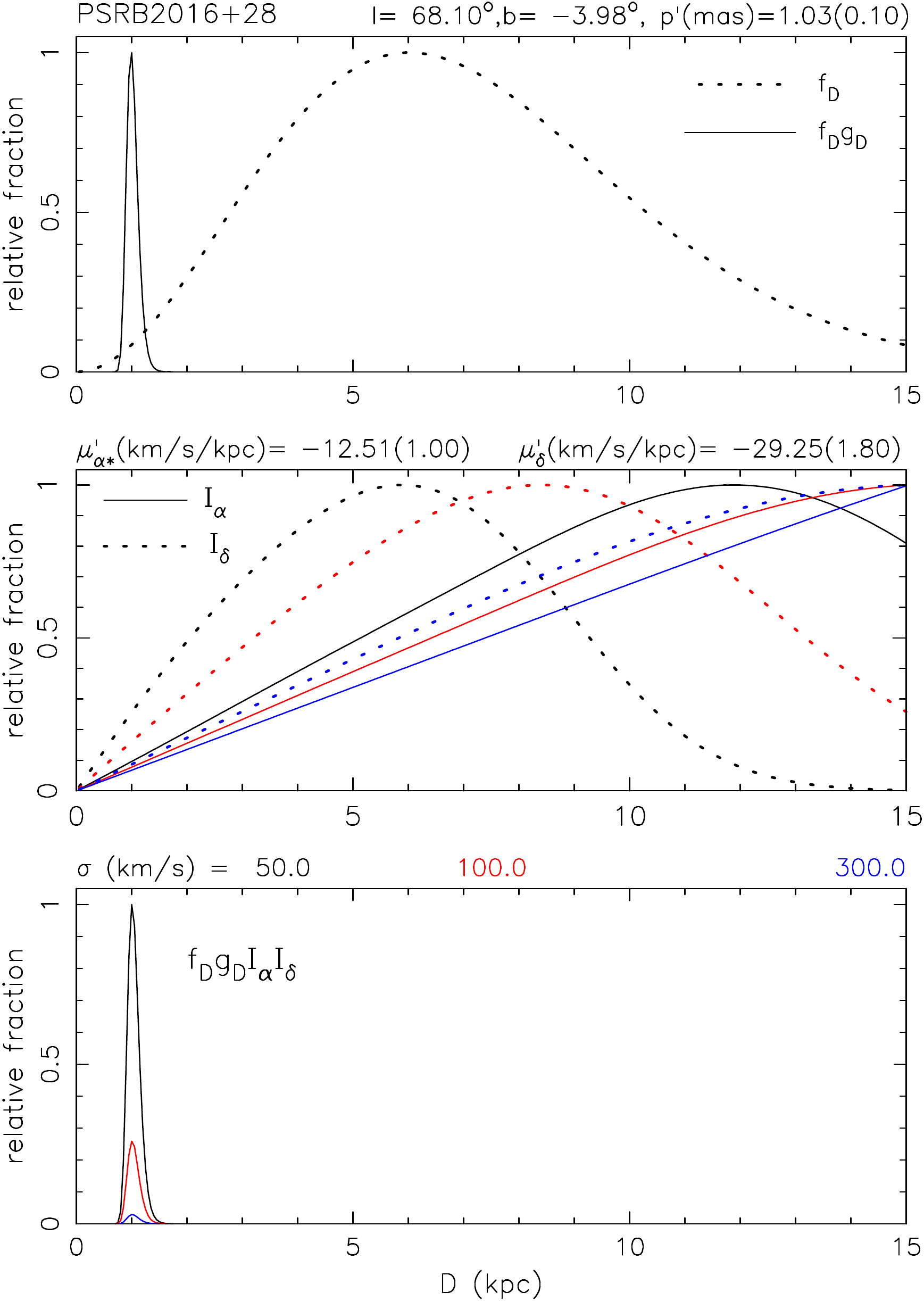}
\caption{The integrand of Eq.\,\ref{e:lmaxw}, as Figure\,\ref{f:lmaxw},
  but now for PSR\,B2016+28. In this case the parallax and galactic
  pulsars distribution (top frame) and proper motion (middle frame)
 imply a small projected velocity $v'_\perp$. This favours the Maxwellian with
 low scaling parameter $\sigma=50$\,km/s. }
\label{f:lmaxwb}
\end{figure}

\subsection{Description with two Maxwellians}

As argued in Section\,\ref{s:vel}, a single Maxwellian is not a
good description of the observed velocity distribution. To
illustrate this,we show in Figure\,\ref{f:lmaxwb} that the data for
PSR B2016+28 (taken from Brisken et al.\, 2002) imply
a low projected velocity: $v'_\perp=31$\,km/s. From the three
Maxwellians considered, this velocity clearly favours the one
with $\sigma=50$\,km/s.

\begin{figure}[t]
\includegraphics[width=\columnwidth]{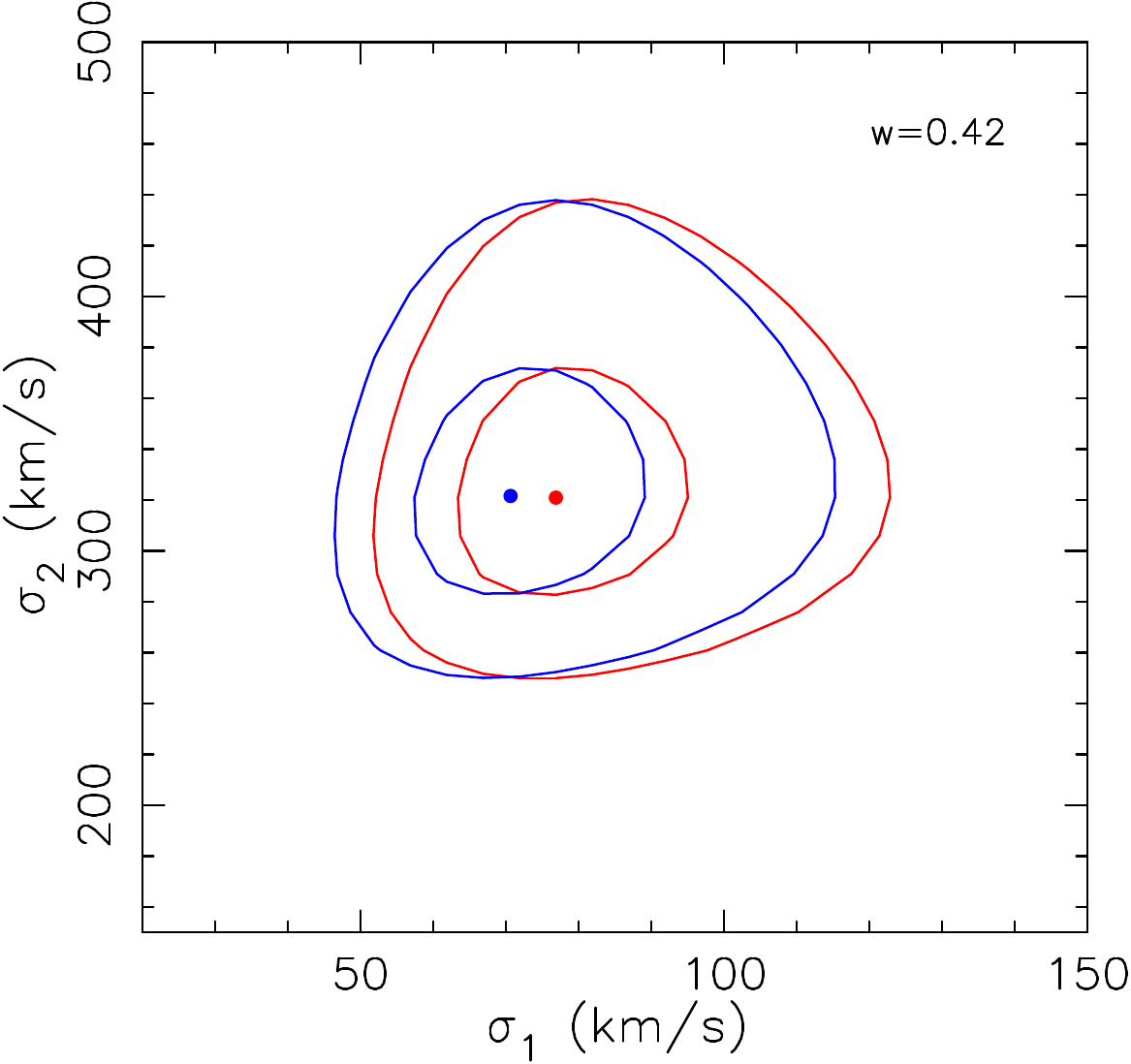}
\caption{Red: Contours indicating the allowable range of the best
solution for two Maxwellians, for $w=0.42$. The best
solution is given as a point, the contours contain 68\%\ and 95\%\ probability
($\Delta\mathcal{L}_{2\mathrm{maxw}}=1$ and
$\Delta\mathcal{L}_{2\mathrm{maxw}}=4$, respectively). Blue: the same
for a model in which the corrections for galactic rotation are omitted.}
\label{f:l2maxw}
\end{figure}

As a second approach to the determination of the intrinsic
velocity distribution of young pulsars, we therefore describe
it with the sum of two Maxwellians:
\begin{equation}
f_v(\vec\sigma) = \sqrt {\frac{2}{\pi}} v^2
\left[\frac{w}{\sigma_1^3} \exp \left( -\frac{1}{2}
    \frac{v^2}{\sigma_1^2} \right) 
 + \frac{(1-w)}{\sigma_2^3}  \exp \left( -\frac{1}{2} \frac{v^2}{\sigma_2^2} \right) \right]
\label{e:twomaxw}\end{equation}
with the parameter vector $\vec\sigma=w,\sigma_1,\sigma_2$.
In analogy we Eqs.\,\ref{e:lmaxw}, \ref{e:total}, \ref{e:callik} we now have
\begin{equation}
L_{2\mathrm{maxw}} (\vec\sigma) = w L_\mathrm{maxw} (\sigma_1) + (1-w) L_\mathrm{maxw} (\sigma_2)
\label{e:l2maxw}
\end{equation}
\begin{equation}
\mathcal{L}_{2\mathrm{maxw}} (\vec\sigma) = -2\sum_{i=1}^N\ln
L_{2\mathrm{maxw},i} (\vec\sigma)
\label{e:2max_lik}
\end{equation}
\begin{equation}
\Delta\mathcal{L}_{2\mathrm{maxw}}(\vec\sigma)\equiv \mathcal{L}_{2\mathrm{maxw}}(\vec\sigma) -
\mathcal{L}_{2\mathrm{maxw}}(\vec\sigma_\mathrm{opt})
\label{e:callik2}\end{equation}
Verbunt et al. (2017) compute $L_\mathrm{maxw}(\sigma)$ on a grid of $\sigma$ values with
a spacing of 1\,km/s, and use the {\tt amoeba} routine from Press et
al.\ (1986) to determine the values of $\vec\sigma_\mathrm{opt}$ that minimize
$\mathcal{L}_{2\mathrm{maxw}}$. They find that the best description of
the velocity distribution is the combination of 42\%\ of the pulsars
in  a Maxwellian with $\sigma_1=77$\,km/s with a 58\%\
in a Maxwellian with $\sigma_2=320$\,km/s.
Comparing the best solution for two Maxwell- ians with that for one
Maxwellian, Verbunt et al.\ find $\mathcal{L}_{2\mathrm{maxw}}(\vec\sigma_\mathrm{opt})-
\mathcal{L}_\mathrm{maxw}(\sigma_\mathrm{opt})=-14$. For two added
parameters this difference indicates that the solution with two
Maxwellians is significantly better.

The choice of $\mathcal{L}_\mathrm{2maxw}$ according to
Eq.\,\ref{e:l2maxw} implies that the distribution of 
$\Delta\mathcal{L}_{2\mathrm{maxw}}(\vec\sigma)$ approximates a
$\chi^2$ distribution. Thus we draw find the 68\%\ and 95\%\
probability contours in the $\sigma_1$ - $\sigma_2$ plane 
as delineated by 
$\Delta\mathcal{L}_{2\mathrm{maxw}}(\vec\sigma)=1$ and 
$\Delta\mathcal{L}_{2\mathrm{maxw}}(\vec\sigma)=4$,
respectively.
This is shown in Figure\,\ref{f:l2maxw}.

To gauge the effect of the corrections for galactic rotation, we show
in the same Figure the results for a computation in which these
corrections were set to zero. This leads to a marginal shift to a 
lower value (71\,km/s) for $\sigma_1$. The value of
$\sigma_2$ is not affected.

\section{Conclusions}

The distance derived from a parallax measurement of a single pulsar
is subject to bias, because the distance prior of pulsars  is
not constant. Application to pulsar \psrj\ of the correct method
for a realistic spatial distribution of millisecond pulsars
shows that the isotropic gamma-ray flux of
this recycled pulsar is more than 10\%\ of its spindown luminosity.
 
For the determination of spatial velocities of young, in the
sense of not recycled, pulsars we only have measurements of the
projections $v_\perp$ of these velocities on the celestial sphere.
The most direct measurements of $v_\perp$ are obtained from
VLBI observations of parallax and proper motion.
Timing observations can also be used, but the measurement
uncertainties are
generally several orders of magnitude larger, allowing for
determinations of proper motions, but only giving upper limits
to the parallaxes.
Indirect measurements of distances from disperions measures
depend on models for the electron distribution in the Milky Way,
and as a result the uncertainties in the distances thus derived are
large, and not gaussian but systematic.

Detailed analysis of the parallaxes and proper motions of 28 pulsars
confirms the suspicion based on a rough analysis that a single
Maxwellian does not describe the velocity distribution of these
pulsars. A description with two Maxwellians is significantly better,
and finds as a best solution that 42\%\ of the pulsars follow a
Maxwellian with distribution parameter $\sigma_1=77$\,km/s, and 58\%\
a Maxwellian with $\sigma_2=320$\,km/s.
This detailed analysis considers pulsar velocities with respect to their
local standard of rest, and to do so applies corrections for
galactic rotation. At the current level of accuracy, however, it
turns out that these corrections do not have a significant impact
on the result.

The number of 28 pulsars for which accurate measurements are
availabe is too small to conclude that the velocity distribution
is indeed given by the sum of two Maxwellians. It is clear
that pulsars have a wide range of velocities, but to determine the
exact form of the distribution, accurate measurements of more
pulsars are necessary.


\appendix

\section{The Maxwellian velocity distribution and its projection}
The Maxwellian velocity distribution may be written
\begin{equation}
f(v)dv =\sqrt{2\over\pi}{v^2\over\sigma^3}  e^{-v^2/(2\sigma^2)}dv
\end{equation}
In the isotropic case, the Maxwellian can be decomposed in three 
gaussian distributions with the same $\sigma$ but otherwise
independent, along three mutually perpendicular directions.
In the $x$-direction, for example, we have
\begin{equation}
f(v_x)dv_x = {1\over\sigma\sqrt{2\pi}} e^{-{v_x}^2/(2\sigma^2)}dv_x
\end{equation}
Choosing the $z$-direction along the line of sight, we  find for
the velocity perpendicular to the line of sight
\begin{eqnarray}
f(v_\perp)dv_\perp &=&
{1\over2\pi\sigma^2}e^{-(v_x^2+v_y^2)/(2\sigma^2)}dv_xdv_y\nonumber \\
& = &
{1\over\sigma^2}e^{-v_\perp^2/(2\sigma^2)}v_\perp dv_\perp
\end{eqnarray}
The cumulative distribution of $v_\perp$ follows as
\begin{eqnarray}
f(v_\perp<v_c) &=&
\int_0^{v_c} {1\over\sigma^2}e^{-v_\perp^2/(2\sigma^2)}v_\perp
dv_\perp\nonumber \\
&=&  1 - e^{{-v_c}^2/(2\sigma^2)} \label{e:vperpcum}
\end{eqnarray}

\section*{Acknowledgement}
We thank Andrei Igoshev for discussions.



\end{document}